\documentclass[aps,twocolumn,prc,superscriptaddress,noshowpacs,nofootinbib,noshowkeys,floatfix]{revtex4}
\usepackage[dvips]{graphics,graphicx}
\usepackage[colorlinks=true,linktocpage=true,linkcolor=blue,citecolor=blue]{hyperref}
\usepackage[usenames,dvipsnames]{color}
\usepackage{amsmath, amssymb}
\usepackage{multirow}
\usepackage{longtable}
\usepackage{color}
\usepackage{xcolor}
\usepackage{xspace}
\usepackage{cleveref}
\usepackage[normalem]{ulem}  

\newcommand{\pt}{$p_{\rm{T}}$\xspace}
\newcommand{\hf}{heavy-flavour\xspace}
\newcommand{\hfs}{heavy-flavours\xspace}
\newcommand{\pta}{$p_{\rm{T}}^{\rm assoc}$\xspace}
\newcommand{\ptt}{$p_{\rm{T}}^{\rm trig}$\xspace}
\newcommand{\ptd}{$p_{\rm{T}}^{\rm D}$\xspace}

\newcommand{\delphi}{$\Delta \varphi$\xspace}

\newcommand{\bef}{\begin{figure}}
\newcommand{\eef}{\end{figure}}
\newcommand{\bc}{\begin{center}}
\newcommand{\ec}{\end{center}}

\newcommand{\be}{\begin{equation}}
\newcommand{\ee}{\end{equation}}
\newcommand{\bea}{\begin{eqnarray}}
\newcommand{\eea}{\end{eqnarray}}

\newcommand{\dph}{\Delta\varphi }

\begin{document}

\title{{\Large Jet fragmentation via azimuthal angular correlations of heavy-flavours in pp collisions at $\sqrt{s} = $ 7 TeV}}

\author{Ravindra Singh,\footnote{Corresponding author.}}
\email{ravirathore.physics@gmail.com}
\author{Swapnesh Khade}
\author{Ankhi Roy}
\email{ankhi@iiti.ac.in}
\affiliation{Department of Physics, School of Basic Sciences, Indian Institute of Technology Indore, Simrol, Indore 453552, India}

\begin{abstract}

Measurements in heavy-flavour azimuthal angular correlation provide insight into the production, propagation, and hadronization of heavy-flavour jets in ultra-relativistic hadronic and heavy-ion collisions. These measurements across different particle species help to isolate the possible modification in particle production and fragmentation due to different mass and quark contents. Jet correlation studies give direct access to the initial parton dynamics produced in these collisions. \\
This article studies the azimuthal angular correlations of heavy-flavour hadrons (charm and beauty mesons and charm baryons) in pp collisions at $\sqrt{s} =$ 7 TeV using PYTHIA8. We study the production of heavy-flavour jets with different parton-level processes, including multi-parton interactions and different color reconnection prescriptions. The heavy-flavour hadrons correlations are calculated in the different triggers and associated \pt intervals to characterize the impact of hard and soft scattering. The yields and the widths associated with the near-side (NS) and away-side (AS) correlation peaks are calculated and studied as a function of associated \pt for different trigger \pt ranges. 

\end{abstract}
\date{\today}
\maketitle

\section{Introduction}
\label{intro}

The Relativistic Heavy-Ion Collider (RHIC) at BNL, USA, and the Large Hadron Collider (LHC) at CERN, Geneva, Switzerland, serve the purpose of studying the exotic state of matter like Quark-Gluon Plasma (QGP)~\cite{STAR:2005gfr,Braun-Munzinger:2015hba}, and unravel its properties by colliding high energy nuclei. These collider experiments aim to probe the strongly interacting matter phase diagram based on Quantum Chromodynamics (QCD). Ultra-relativistic heavy-ion and hadronic (pp) collisions result in the formation of a dense system composed of low transverse momentum (\pt) partons~\cite{ALICE:2016fzo}. Initial hard scatterings in pp, p–Pb, and Pb–Pb collisions produce heavy-flavours, namely charm (c) and beauty (b)~\cite{Braun-Munzinger:2003pwq,Frixione:1997ma,Norrbin:2000zc,Alberico:2013bza,Levai:2009mn}. Their early production can be attributed to their large mass, which allows them to traverse through the QGP and interact with the partons of hot medium. The production cross-section of these heavy quarks is usually calculated using the factorization theorem
\begin{align}
d\sigma_{AB\rightarrow C}^{\rm{hard}} = \Sigma_{a,b,X} f_{a/A} (x_a,Q^2) \otimes f_{b/B} (x_b,Q^2) \otimes\\
\nonumber d\sigma_{ab\rightarrow cX}^{\rm{hard}} (x_a,x_b,Q^2) \otimes D_{c\rightarrow C}(z,Q^2)
\end{align}

where, $f_{a/A} (x_a,Q^2)$ and $f_{b/B} (x_b,Q^2)$ are the parton distribution functions which give the probability of finding parton "a"(b) inside the particle "A"(B) for given x (fraction of particle momentum taken by parton) and factorization scale ($Q^2$), $d\sigma_{ab\rightarrow cX}^{\rm{hard}} (x_a,x_b,Q^2)$ is the partonic hard scattering cross-section, and $D_{c\rightarrow C}(z,Q^2)$ is the fragmentation function of the produced parton (particle). This leads to universal hadronization, but new PYTHIA8 tunes have incorporated different hadronization models based on beyond-leading color approximation (BLC tunes) and rope hadronization (Shoving) which do not assume universal hadronization. The high momentum (\pt) partons through fragmentation (parton showering) \cite{ALICE:2019oyn,ALICE:2021kpy,CMS:2020geg} and hadronization form a cluster of final state particles known as a jet. The study of high-pT jets reveals how parton fragments into various particles and allows the study of the parton's interaction with the medium.

One of the methods to study interactions of heavy-flavours with partons of hot QCD matter is two-particle angular correlation function~\cite{Beraudo:2014boa,PHENIX:2018wex,Zhang:2019bkf,Zhang:2018ucx}, i.e. the distribution of the differences in azimuthal angles, $\Delta \varphi = \varphi_{assoc}-\varphi_{trig}$, and pseudorapidities, $\Delta \eta = \eta_{assoc} - \eta_{trig}$, where $\varphi_{assoc}$ ($\eta_{assoc}$) and $\varphi_{trig}$ ($\eta_{trig}$) are the azimuthal angles (pseudorapidities) of the associated and trigger particles respectively. The structure of the correlation function usually contains a "near side" (NS) peak and an "away side" (AS) peak at $\Delta \varphi=0$ and $\Delta \varphi=\pi$ respectively over a wide range of $\Delta \eta$. In QCD, leading order (LO) heavy-flavour production processes imply back to back correlations at $\Delta \varphi=0$ and $\Delta \varphi=\pi$ with the same distribution parameters, however next-to-leading order (NLO) processes like gluon splitting and flavour excitation can lead to change in the away side peak. Additionally, the production of heavy-flavour hadrons is sensitive to both the charm and beauty fragmentation functions as well as the hadronization mechanisms; for these reasons, the two-particle angular correlation function not only enables us to study how heavy-flavours interact with QGP in Pb--Pb collisions but also to characterize the production, fragmentation, and hadronization of heavy-flavour hadrons in pp collisions~\cite{Norrbin:2000zc}. Apart from above mentioned reasons, modification of the correlation function is also possible in the case of p--Pb due to cold-nuclear matter effects (nuclear shadowing and gluon saturation)~\cite{PHENIX:2010hmo,Eskola:2001gt,Kharzeev:2005zr}. After measuring the nuclear modification factor of D mesons and electrons from \hf hadron decay in p--Pb collisions at $\sqrt{s_{NN}} = 5.02$ TeV, a small influence of cold-nuclear matter effects on \hf quark production at midrapidity was observed~\cite{CMS:2017qjw,ALICE:2015zhm,ALICE:2018lyv,STAR:2014wif,Fujii:2013yja}.

In this article, we present the study of the azimuthal correlation function of prompt D mesons/baryons and B mesons with charged hadrons in pp collisions at $\sqrt{s}$ = 7 TeV using PYTHIA8, where "prompt" refers to D mesons produced from the fragmentation of charm-quark generated in initial hard scattering, including those from the decay of excited charmed resonances and excluding D mesons produced from beauty hadron weak decays. In terms of particle multiplicity and angular profile, the near-side correlation peak is a suitable probe for characterizing charm jets and their internal structure. Probing the near-side peak~\cite{ATLAS:2012cix} features as a function of charged-particle transverse momentum (\pt), possibly up to values of a few GeV/c, can provide insight into the transverse-momentum distribution of the jet constituents. These features are useful to decifer how the jet momentum fraction not carried by the D mesons is shared among the other particles produced by charm fragmentation, as well as the correlation between the \pt of these particles and their radial displacement from the jet axis. Variations in the amplitude and width of the away-side peak also shed light on the dynamics of \hf production mechanism~\cite{Albacete:2018ruq}.\\
Various event generators in high energy physics mainly use either string model or cluster model for the description of hadronization~\cite{Andersson:1983jt,Hwa:1979pn,Buckley:2011ms}. This study aims to understand and compare the fragmentation and hadronization of D mesons/baryons and B mesons using different tunes of PYTHIA8. In PYTHIA8, the LUND string hadronization model with parameters tuned using $e^+e^-$ data is used for the fragmentation process~\cite{Sjostrand:2006za,Sjostrand:2007gs,Campbell:2022qmc,Buckley:2011ms}. Different tunes of PYTHIA8 such as Monash, 4C, Mode(0,2,3), and shoving differ in implementations of string hadronization which are discussed in the next section. The production and the fragmentation of charmed baryons and beauty mesons is inherently different owing to the difference in their quark content. It will be interesting not only to see which of these models gives a better description of charmed mesons data but also their predictions for charmed baryons and beauty mesons. In the literature, the hadronization of these particles is also explained by $3 \rightarrow 1$ and $2 \rightarrow 1$ coalescence model~\cite{Biro:1994mp,Fries:2008hs}. As far as the comparison between charmed mesons and beauty mesons is concerned, global fragmentation functions based on Next to Leading Logarithmic (NLL) calculations contain the parameter which is a function of the inverse square of heavy-flavour mass~\cite{Kniehl:2007erq,Salajegheh:2019ach,Epele:2018ewr,Kramer:2018vde,Kramer:2017gct}. We anticipate that the effect of mass hierarchy between charm and beauty quark should also be visible in azimuthal angular correlation.

The paper is organized as follows, In section~\ref{evt_gen}, we discuss the event generation and analysis methodology, followed by the results of our analysis in section~\ref{result}, then we summarise our findings in section~\ref{sum}.

\section{Event generation and Analysis methodology}
\label{evt_gen}

\begin{figure*}[ht!]
    \centering
    \includegraphics[scale = 0.8]{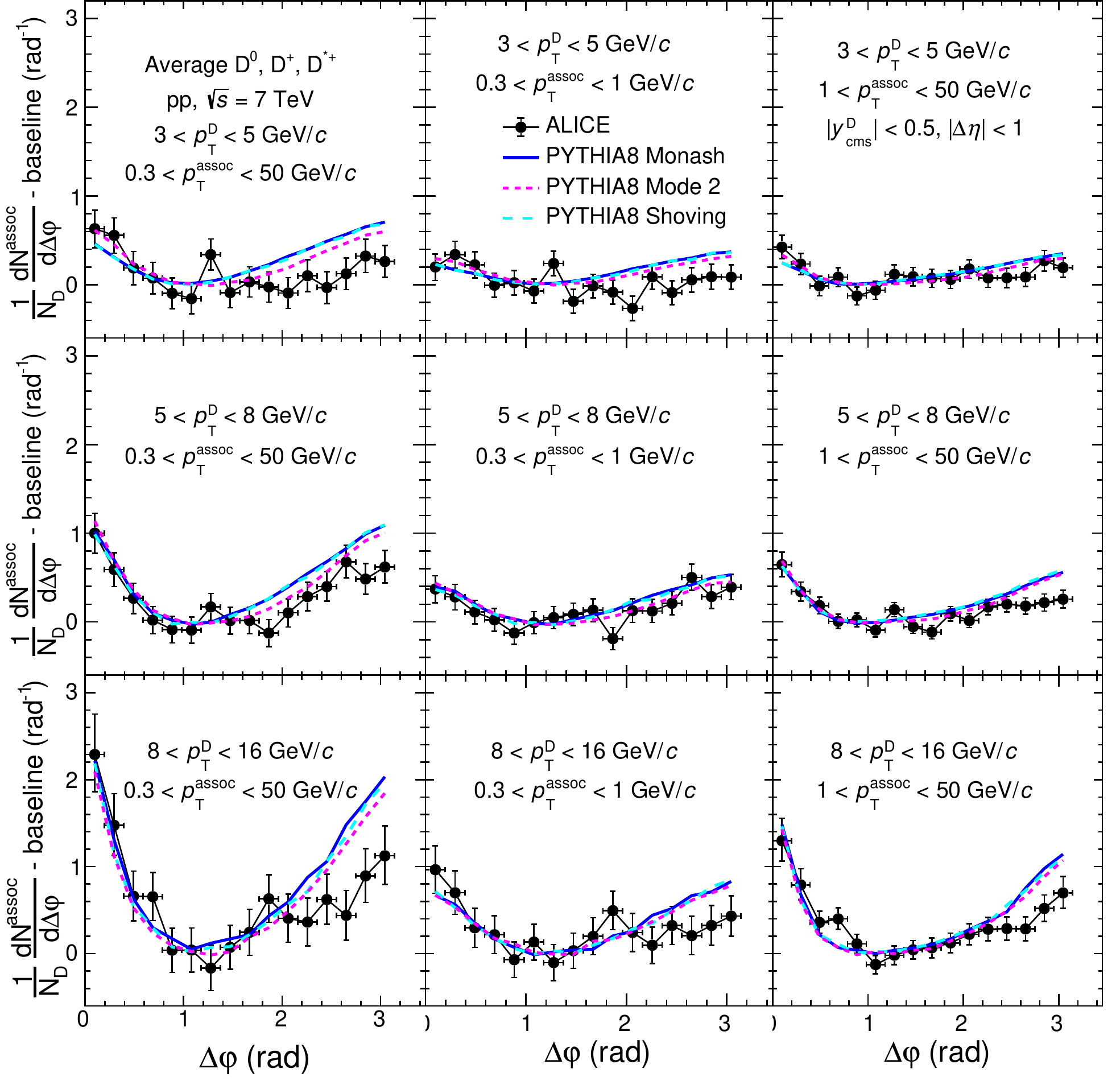}
    \caption{Comparison of ALICE results of average D meson azimuthal-correlation distribution with PYTHIA8 (Monash, Mode 2, and Shoving) after baseline subtraction for $3 < p_{\rm{T}}^{\rm D} < 16$~GeV$/c$ and for different associated \pta ranges in pp collisions at $\sqrt{s} = 7$ TeV.}
    \label{fig:delphi1}
\end{figure*}

\begin{figure*}[ht!]
    \centering
    \includegraphics[scale = 0.8]{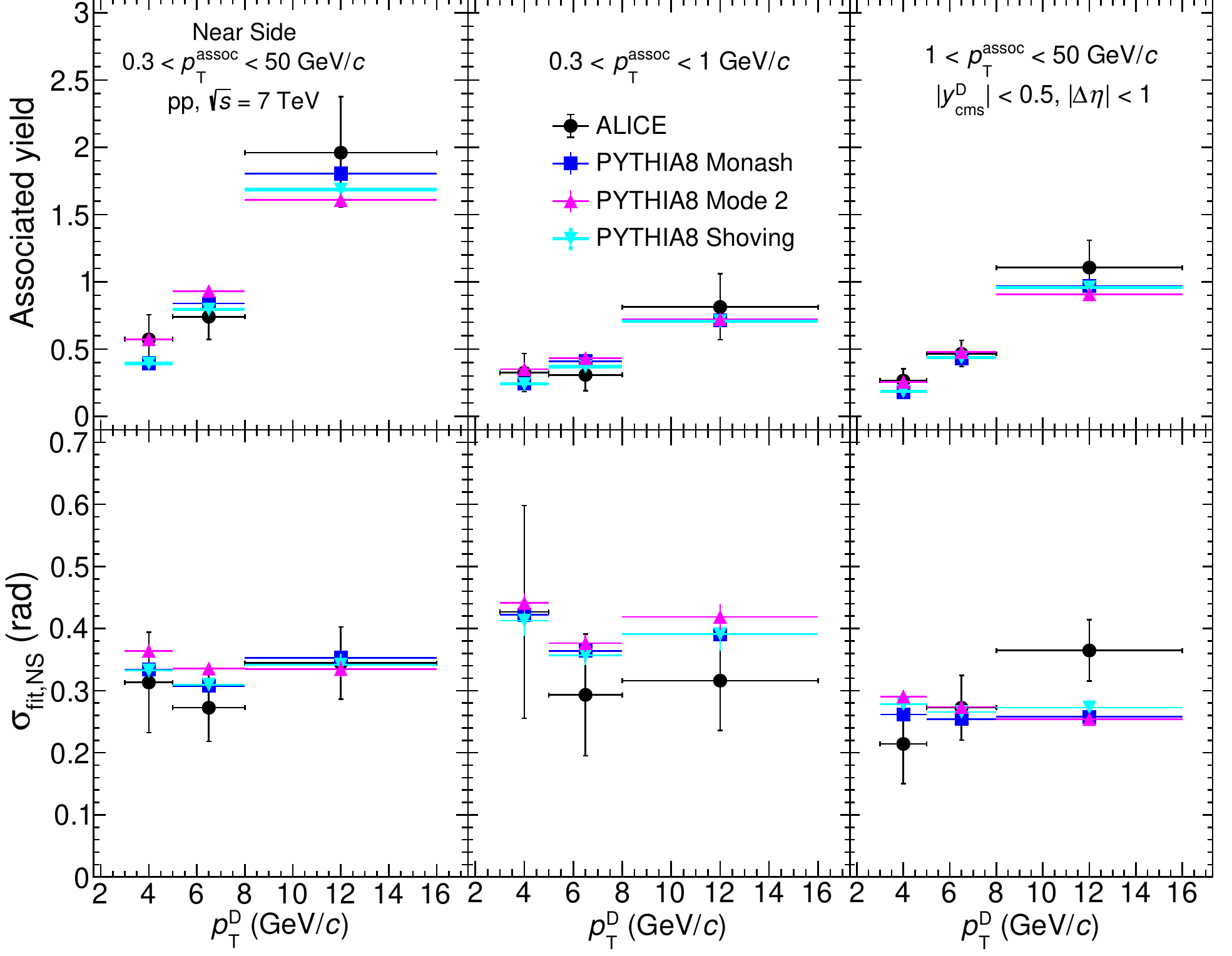}
    \caption{Comparison of ALICE result of average D meson near-side yields (top) and widths ($\sigma$) with PYTHIA8 (Monash, Mode 2, and Shoving) in pp collisions at $\sqrt{s} = 7$ TeV for $3 < p_{\rm{T}}^{\rm D} < 16$~GeV$/c$ in different associated \pta ranges.}
    \label{fig:Dyieldsigma_with ALICE}
\end{figure*}

\begin{figure*}[ht!]
    \centering
    \includegraphics[scale = 0.8]{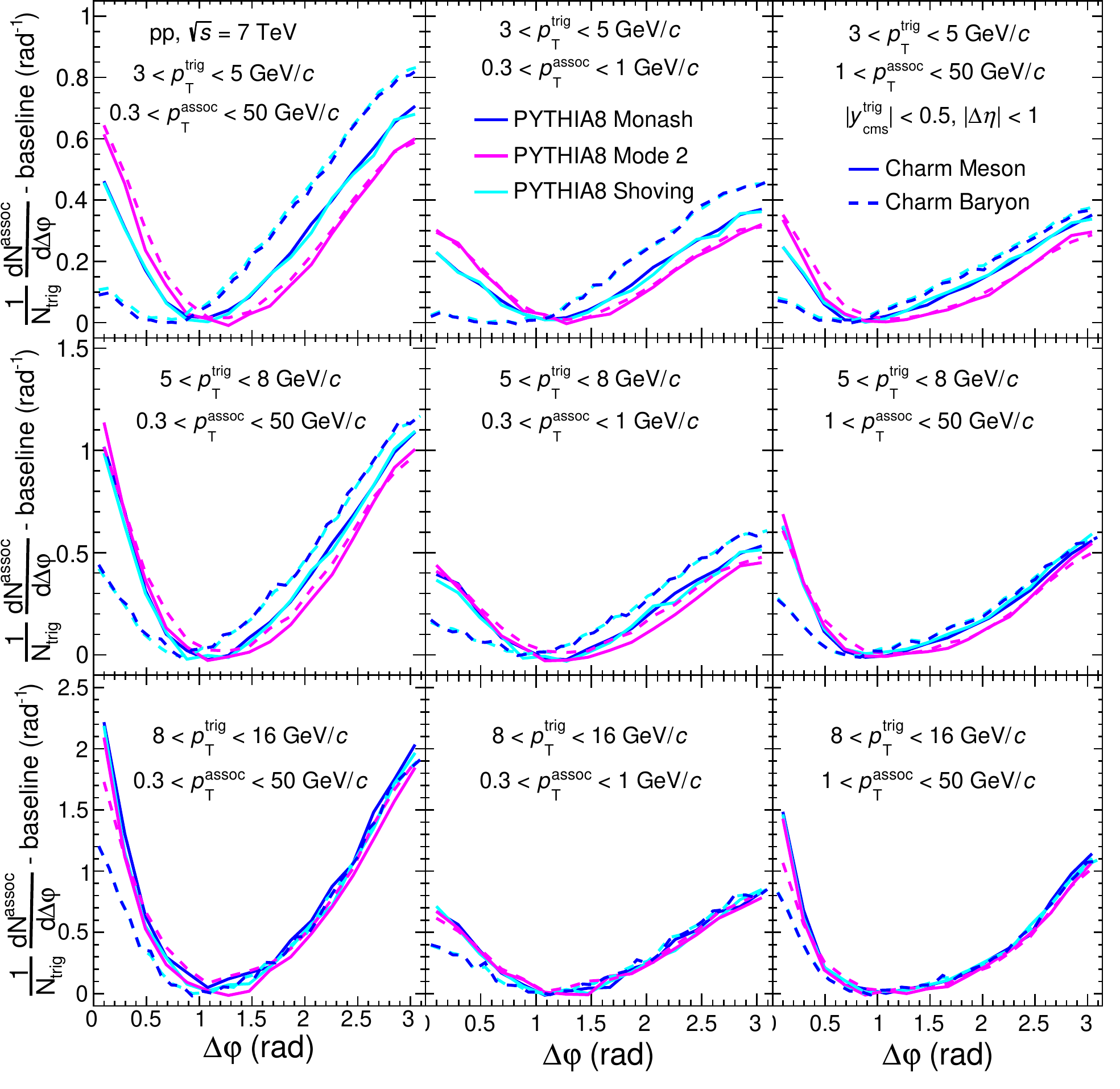}
    \caption{Comparison of average charmed meson and baryon azimuthal-correlation distribution derived from PYTHIA8 (Monash, Mode 2, and Shoving) after baseline subtraction for $3 < p_{\rm{T}}^{\rm trig} < 16$~GeV$/c$ and for different associated \pta ranges in pp collisions at $\sqrt{s} = 7$ TeV.}
    \label{fig:MBDelphi}
\end{figure*}

\begin{figure*}[ht!]
    \centering
    \includegraphics[scale = 0.8]{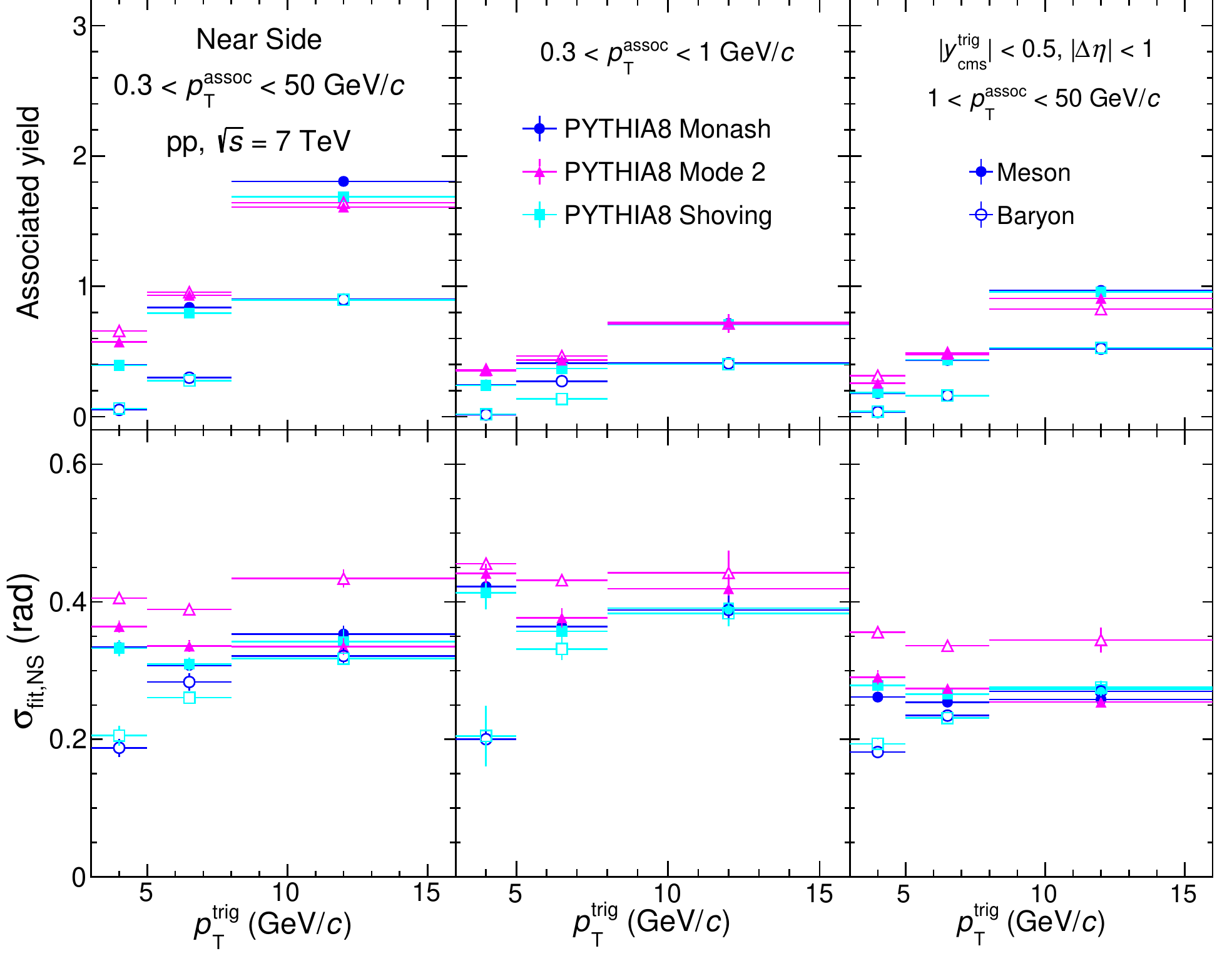}
    \caption{Comparison of average charmed meson and baryon near-side yields and widths ($\sigma$), derived from PYTHIA8 (Monash, Mode 2, and Shoving) after baseline subtraction in pp collisions at $\sqrt{s} = 7$ TeV for $3 < p_{\rm{T}}^{\rm D} < 16$~GeV$/c$ in different associated \pta ranges.}
    \label{fig:MByieldsigma}
\end{figure*}

\begin{figure*}[ht!]
    \centering
    \includegraphics[scale = 0.8]{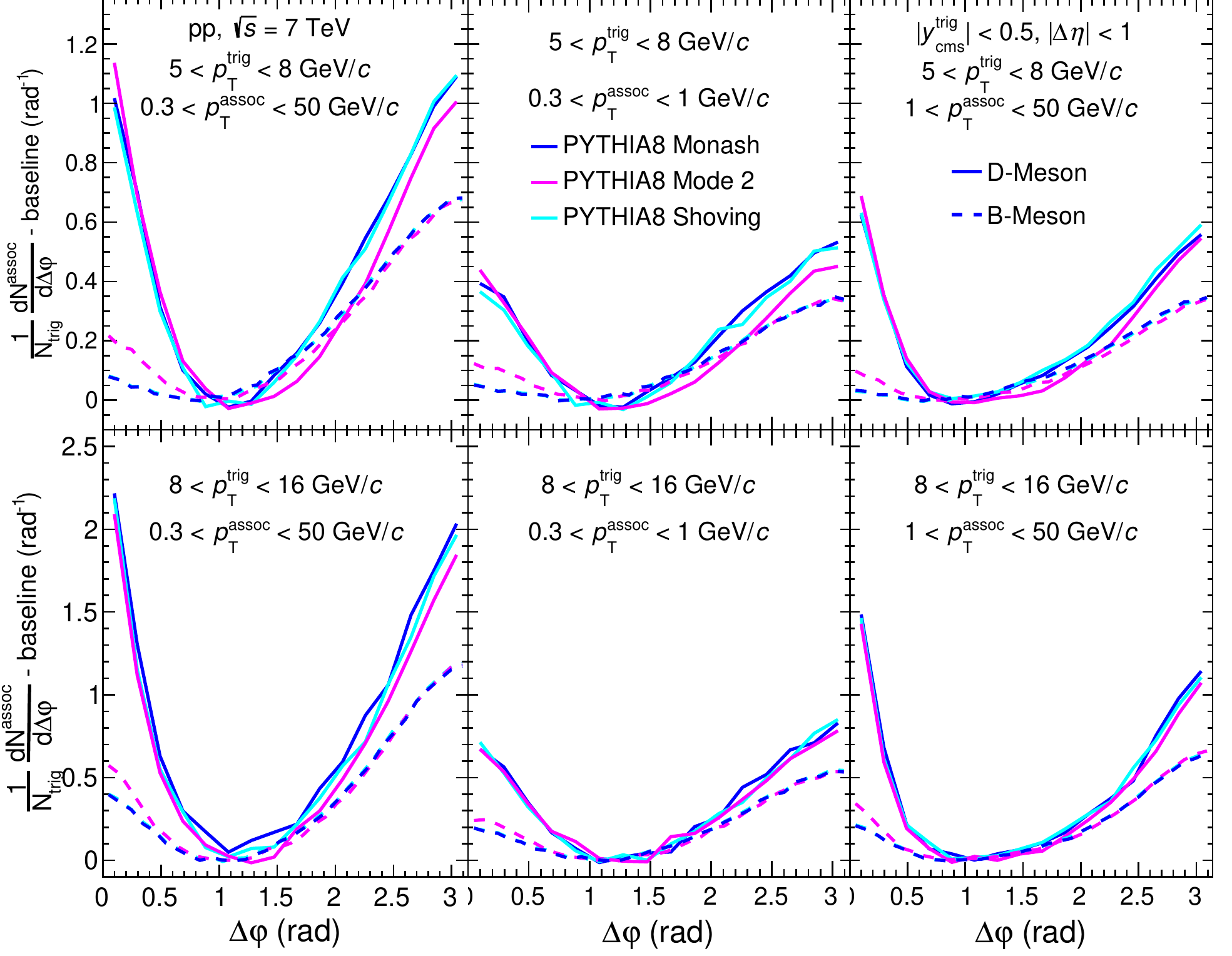}
    \caption{Comparison of average charm and beauty meson azimuthal-correlation distribution derived from PYTHIA8 (Monash, Mode 2, and Shoving) after baseline subtraction for $5 < p_{\rm{T}}^{\rm trig} < 16$~GeV$/c$ in different associated \pta ranges in pp collisions at $\sqrt{s} = 7$ TeV.}
    \label{fig:DBdelphi}
\end{figure*}

\begin{figure*}[ht!]
    \centering
    \includegraphics[scale = 0.8]{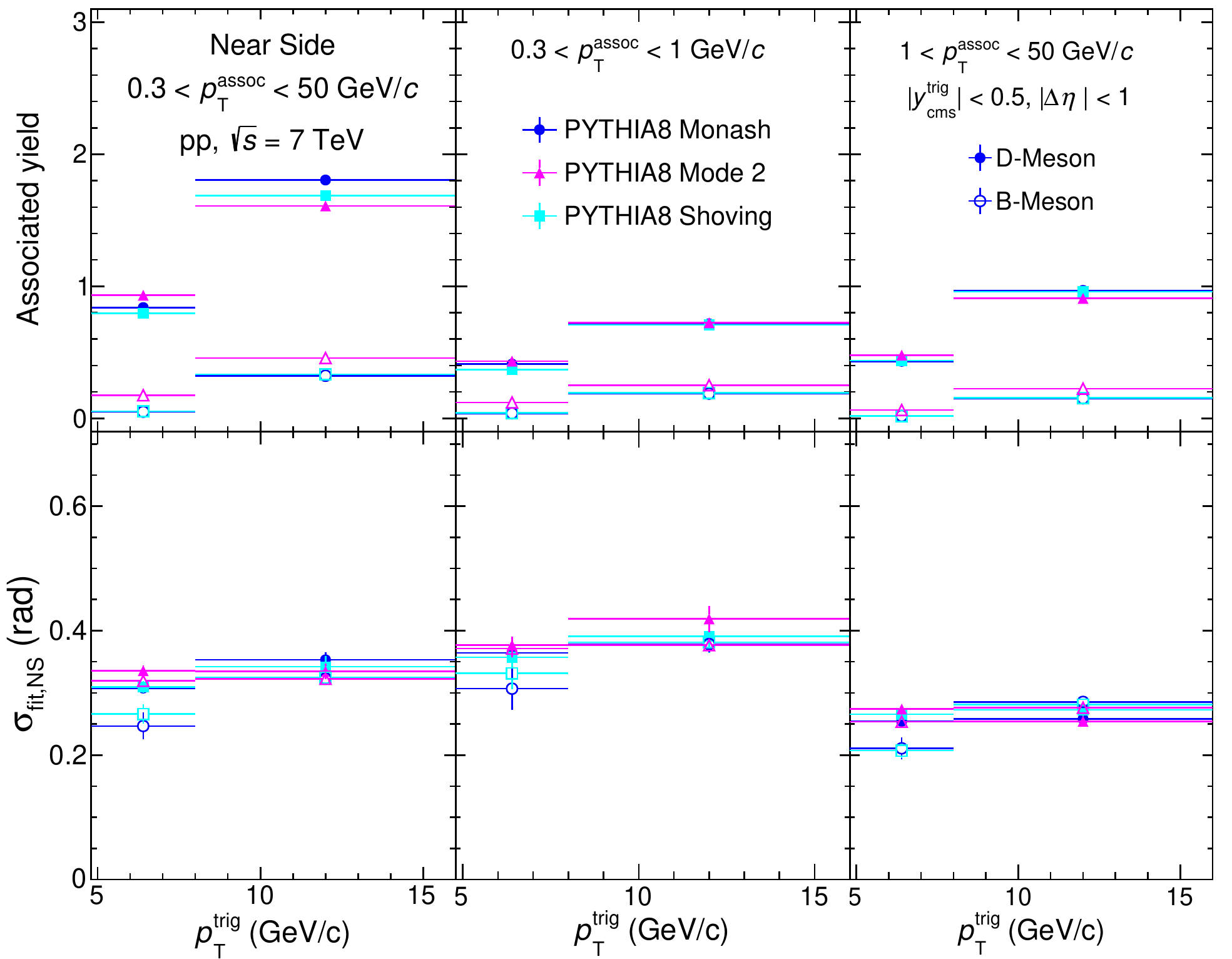}
    \caption{Comparison of average charm and beauty meson yields and widths ($\sigma$) derived from PYTHIA8 (Monash, Mode 2, and Shoving) after baseline subtraction in pp collisions at $\sqrt{s} = 7$ TeV for $5 < p_{\rm{T}}^{\rm trig} < 16$~GeV$/c$ in different associated \pta ranges.}
    \label{fig:DByieldsigma}
\end{figure*}

The PYTHIA event generator~\cite{Sjostrand:2006za,Singh:2021edu,Sjostrand:2007gs,Bierlich:2018xfw} is used to investigate proton-proton, proton-lepton, proton-nucleus, and nucleus-nucleus collisions in depth. It employs 2 → 2 QCD matrix elements evaluated perturbatively with leading-order precision, with the next-to-leading order contributions taken into account during the parton showering stage. The parton showering follows a leading-logarithmic \pt ordering, with soft-gluon emission divergences excluded by an additional veto, and the hadronization is handled with the Lund string-fragmentation model. It offers a plethora of processes and tunes from which to choose and apply based on the physics involved in the study. PYTHIA employs multi-parton interactions (MPI)~\cite{Corke:2009tk,Diehl:2011yj,Sjostrand:2017cdm} with incoming parton beams, employing hard and soft scattering processes followed by Initial-State Radiations (ISR) and Final-State Radiations (FSR). The high \pt partons give rise to showers or jets that fragment and hadronize according to the Lund string fragmentation model~\cite{Andersson:1983ia}. Hadronization is accomplished through the Color Reconnection (CR) mechanism between partons~\cite{Khoze:1994fu,Gieseke:2012ft,Rathsman:1998tp}, which is accomplished by rearranging the strings between them. This modifies the total string length, which affects the hadronization process. When the string length is small enough after the subsequent creation of light quark-antiquark pairs, the partons hadronize to a hadron. The MPI and CR phenomena in PYTHIA play an essential role in the particle production mechanism, as evidenced by the charged-particle multiplicity distributions.

The CR mechanism of hadronization can be investigated further by looking at the string topology between the partons. The Leading Color (LC) approximation assigns a unique index to quarks and antiquarks connected by a colored string. This guarantees a fixed number of colored strings, ensuring that no two quarks (antiquarks) have the same color. The same is true for gluons, which are represented by a pair of colored quark-antiquark. This model is extended to non-LC topologies, also known as Beyond-LC (BLC)~\cite{Christiansen:2015yqa}, in which colored strings can form between LC and non-LC connected partons. This opened the possibility of a string being linked to partons of matching indexes other than the LC parton. Three modes of Color Reconnection in the BLC approximation are used with the different constraints on the allowed string reconnections, taking into account causal connections of dipoles involved in a reconnection and time dilation effects caused by relative boosts between string pieces~\cite{Skands:2014pea,Christiansen:2015yqa}. We investigated different PYTHIA8/Angantyr tunes, i.e., LC (MONASH 2013~\cite{Skands:2014pea}, and 4C~\cite{Corke:2010yf} ), BLC (Mode0, Mode2, Mode3), and rope hadronization (Shoving)~\cite{Bierlich:2016vgw,Bierlich:2017vhg,Bierlich:2014xba,Bierlich:2015rha}. In our study, similar results were obtained with the LC tunes 4C and Monash, and different BLC tunes were also consistent with one another; therefore, for this investigation, we used the Monash, Mode2, and Shoving tunes and investigated how different hadronization processes affected the results. 

Leading order (LO) perturbative scattering processes of gluon fusion ($gg \rightarrow Q\overline{Q}$) or pair annihilation ($q\overline{q} \rightarrow Q\overline{Q}$) is used for the production of \hfs in PYTHIA. PYTHIA also approximates certain higher-order contributions within its LO framework via flavour excitations ($gQ \rightarrow Qg$), or gluon splittings ($g\rightarrow Q\overline{Q}$) which give rise to heavy-flavour production during high \pt parton showers~\cite{Braun-Munzinger:2003pwq,Norrbin:2000zc}.

We used PYTHIA version 8.3 to generate around 1B events for each tune in pp collisions at $\sqrt{s} =$ 7 TeV. heavy-flavour hadrons are selected within $|y| < 0.5$. The \pt of trigger particle (heavy-flavour) is selected in three intervals, i.e., 3-5, 5-8, and 8-16 GeV/$c$, while associate particles are selected in the ranges 0.3-50, 0.3-1, 1-50 GeV/$c$. The inelastic, non-diffractive component of the total cross-section for all soft QCD processes is used with the switch \texttt{SoftQCD:all = on} with MPI. Correlation distribution was obtained by correlating each trigger particle with all the associated charged particles. It is to be noted that the decay product of the trigger particle is excluded from the correlation function. The $\Delta \eta$ is selected in the range from -1 to 1. The correlation distribution is fitted with the generalized Gaussian function for the near-side peak, Gaussian function for the away-side peak, and $0^{th}$ order polynomial the baseline identification as shown in the eq.~\ref{eq:fitfunggaus}.  

\begin{footnotesize}
\begin{equation}
{f(\dph)} = b + \frac{Y_{\rm{NS}}\times\beta_{\rm{NS}}}{2\alpha_{\rm{NS}}\Gamma{(1/\beta_{\rm{NS}})}}\times e^{-(\frac{\dph}{\alpha_{\rm{NS}}})^{\beta_{\rm{NS}}}} + \frac{Y_{\rm{AS}}}{\sqrt{2\pi}\sigma_{\rm{AS}}}\times e^{-(\frac{\dph - \pi}{\sqrt{2}\sigma_{\rm{AS}}})^{2}}
\label{eq:fitfunggaus}
\end{equation}
\end{footnotesize}

Where $Y_{NS}$ and $Y_{AS}$ are the yields for NS and AS peaks, $\beta_{NS}$ is the shape parameter for near-side peak, and $\alpha_{NS}$ is related to the $\sigma_{NS}$ (width) of the peak by the given relation:

\begin{equation}
\sigma_{\rm{NS}} = \alpha_{\rm{NS}}\sqrt{\Gamma(3/\beta_{\rm{NS}})/\Gamma(1/\beta_{\rm{NS}})}    
\end{equation}

In this contribution, we tried to study the fragmentation and hadronization of heavy-flavours via jet-like azimuthal correlation of heavy-flavour hadrons with the charged particle in pp at $\sqrt{s} = 7$ TeV. Charm mesons species which are selected for the comparisons are $\rm {D^{0}}$, $\rm {D^{+}}$ and $\rm {D^{*+}}$, similarly charm baryons species are $\Lambda_{c}^{+}$, $\Sigma_{c}^{0}$, $\Sigma_{c}^{+}$, $\Xi_{c}^{+}$, $\Xi_{c}^{0}$, $\Omega_{c}^{0}$, $\Omega_{c}^{0*}$, and beauty mesons species are $\rm {B^{0}}$, $\rm {B^{+}}$, $\rm {B_{s}^{0}}$ and $\rm {B^{*+}}$ with their anti-particles.


\section{Results} 
\label{result}

The jet-like two-particle correlation measurement is an alternative tool to study the jet properties even at low \pt where direct jet measurement is not possible~\cite{Connors:2017ptx}. The correlation measurements provide insight into particle production from the different processes, i.e., pair creation (LO), gluon-splitting, and flavour-excitation (NLO).

The ALICE measurements of azimuthal correlations for charm mesons are compared with  PYTHIA prediction in the following subsection. The measurements of charm mesons are independently compared to charm baryons and beauty mesons to spot any potential alterations in jet fragmentation. 

\subsection{Comparison with ALICE data}

In order to validate the settings of PYTHIA that are used for this study, the azimuthal correlation between D meson and charged particles from the PYTHIA event generator with different color reconnection (CR) schemes and rope hadronization (RH) model is compared with the measurements of ALICE experiment~\cite{ALICE:2016clc}. In the FIG~\ref{fig:delphi1}, baseline subtracted \delphi distribution compared with ALICE data in triggered D mesons \ptd intervals 3-5, 5-8 and 8-16 GeV/$c$ and associate \pta intervals 0.3-50, 0.3-1, and 1-50 GeV/$c$ in the rapidity range $|y^{D}_{cms}| <$ 0.5. Most of the fraction in the baseline is contributed by the underlying event and dominated by low \pt particles. The qualitative shape of the correlation function and the evolution of the near- and away-side peaks with trigger and associated particle \pt are consistent with ALICE measurement. However, PYTHIA measurements overestimate the away-side peak, especially at high \ptd. This study suggests that PYTHIA needs to reform the fragmentation of particles produced at the recoiling jet. All the tunes of PYTHIA provide the same results for D meson and charged particle correlation. It is observed that the height of the correlation peak is increasing with \ptd, which suggests the production of a higher number of particles in the jet accompanying the fragmenting charm quark when the energy of the trigger particle increases. However, no significant difference was observed among different CR and RH tunes in D mesons correlation measurements.

A more quantitative comparison of the near- and away-side peak features and the \pt evolution can be made by measuring the yields and widths of the peaks. The yields and widths are obtained by fitting with the generalized Gaussian function as discussed in section~\ref{evt_gen}. Yield and width ($\sigma$) of the near-side peaks of D meson and charged particles correlation are shown in FIG~\ref{fig:Dyieldsigma_with ALICE} with different tunes and compared with ALICE results. The peak's yield is shown in the top panel, whereas widths are shown in the bottom panel. The per trigger associated yields of the peak are increasing with increasing trigger particle \ptd. This is expected, as high energetic particles are in general produced by high energetic partons, which in turn fragment into a more significant number of particles. Furthermore, as \pta increases, the associated yield decreases. This is because heavy flavor quarks occupy a larger portion of the phase space during fragmentation.  Hence, the remaining phase space for emitting further high \pt particles is limited, and most of the accompanying associated particles are softer. The near-side peak width ($\sigma$) is shown in the bottom panel of FIG~\ref{fig:Dyieldsigma_with ALICE}. The widths estimated by PYTHIA and from the ALICE measurement are almost flat and consistent with each other within statistical uncertainty.

\subsection{Comparison with charm baryons}

Currently, statistics are not enough to measure the azimuthal correlation of charm baryons experimentally. However, it may be feasible in the upcoming LHC run 3. In the FIG~\ref{fig:MBDelphi}, we attempt to provide a prediction for charm baryons fragmentation and modification of fragmentation compared to charm mesons. It is observed that the height of the near-side peaks is largely suppressed for charm baryons, derived by using default tune Monash and rope hadronization Shoving, whereas the height of the away-side peak is increased compared to charm mesons. In mode 2, charm meson and baryon peaks are consistent with each other. 

Similar to the previous section, the near-side observables obtained from fitting are shown in FIG~\ref{fig:MByieldsigma}. It is clearly seen that the associated yield of charm baryons is almost half estimated from Monash and Shoving. In contrast, in mode 2, charm baryons yield is consistent with charm mesons yield. On the other hand, near-side widths from Monash and Shoving are suppressed with respect to mode 2 for baryons at low \ptt, whereas, at higher \ptt, widths are consistent with charm mesons. A higher width of charm baryons can be seen from the Mode 2 tune for all the \ptt and \pta intervals. The trend was very similar to the production cross sections of charm baryons normalized by $\rm D^{0}$ meson, where tune Monash underestimates the ALICE measurement, on the other hand, Mode 2 is in good agreement with the data, especially for ${\Lambda}_{c}$ baryon~\cite{ALICE:2021dhb}. The new CR tunes introduce new color reconnection topologies, including junctions, that enhance baryon production, and charmonia, to a lesser extent. At the same time, multiparton interactions (MPI) are observed in PYTHIA8 to increase the charm quark production significantly. This leads to the modification of the relative abundances of the charm hadron species. The relative baryon enhancement is only observed when the MPI is coupled to a color reconnection mode beyond the leading color approximation. It is observed that for the charm mesons, predictions from the PYTHIA8 generator with the different tunes are reasonably similar.

\subsection{Comparison with beauty mesons}

A similar comparison is made between charm and beauty meson correlation features. The \delphi distribution of charm mesons with charged particles and beauty mesons with charged particles are shown in FIG~\ref{fig:DBdelphi} for \ptt 5-8 and 8-16 GeV/$c$. Here, a comparison between charm and beauty mesons fragmentation for the \ptt 3-5 GeV/$c$ is not shown as the mass of beauty is $\sim$ 5 GeV/$c$, which results in almost a flat near-side peak. The height of the near- and away-side peaks of the correlation function obtained for B mesons are very small compared to D mesons correlation peaks as the available energy of B mesons for fragmentation is small compared to D mesons in the same \pt range. A more quantitative comparison of correlation peaks from D mesons and B mesons fragmentation can be seen in FIG.~\ref{fig:DByieldsigma}. Yields from D mesons are about 4-5 times higher than from B mesons. One of the reasons for the difference in yield can be attributed to the mass hierarchy between charm and beauty quarks, this hierarchy creeps into the global fragmentation function as a factor of an inverse mass square. At higher \ptt, B mesons associated yield increases more rapidly than D mesons. It is also seen that B mesons associated yield for the near-side peak is larger with Mode 2 compared to Shoving and Monash. The widths of correlation peaks are almost flat, and no difference is found between D mesons and B mesons.

\pagebreak
 \section{Summary}
 \label{sum}
 In this work, we attempt to study the heavy-flavour hadrons correlation with the charged particle in pp collisions at $\sqrt{s}$ = 7 TeV using the PYTHIA8 event generator. This paper studies fragmentation via charm mesons, charm baryons, and beauty mesons. The primary observations of this work are summarised below:
 
 \begin{itemize}
     
     \item The near-side correlation distributions and observables of the D mesons derived by PYTHIA are consistent with the ALICE measurements, but PYTHIA needs to reform the physics at away-side observable as it is slightly overestimates.
    
    \item Due to limited phase space, low \pta particles are produced more than high \pta particles, hence for the same \ptt, yield is higher at low \pta.
     
     \item Near-side associated yields to charm baryons are suppressed in Monash and Shoving tune compared to charm mesons yields. However, the difference is negligible in Mode 2. Similar results were observed in the calculation of the charm baryons production cross sections by the ALICE experiment, where the BLC tune mode 2 was in good agreement with the experimental data.  

    \item Near-side yields from D mesons are almost 4-5 times larger than B mesons yield for the same \ptt. A possible reason for this could be the availability of more  energy for D meson fragmentation due to smaller mass.
    
    \item No significant difference is observed in PYTHIA between D and B mesons widths in the same trigger as well as associated $p_{\rm T}$ ranges, i.e.; the dead cone effect has no major impact on the widths of D and B mesons at current precision as they are both heavy particles. However, It will be interesting to see the dead-cone effect in heavy quarks while comparing it with light quarks correlation distribution.

 \end{itemize}

\section{Acknowledgement} 
R.S. acknowledges the financial support (DST/INSPIRE Fellowship/2017/IF170675) by the DST-INSPIRE program of the Government of India. S.K. acknowledges the financial support provided by the Council of Scientific and Industrial Research (CSIR) (File No.
09/1022(0084)/2019-EMR-I), New Delhi. R.S. is also grateful to the authors of PYTHIA8.


\begin{thebibliography}{50}

\bibitem{STAR:2005gfr}
J.~Adams \textit{et al.} [STAR],
Nucl. Phys. A \textbf{757}, 102-183 (2005)
doi:10.1016/j.nuclphysa.2005.03.085
[arXiv:nucl-ex/0501009 [nucl-ex]].

\bibitem{Braun-Munzinger:2015hba}
P.~Braun-Munzinger, V.~Koch, T.~Sch\"afer and J.~Stachel,
Phys. Rept. \textbf{621} (2016), 76-126
doi:10.1016/j.physrep.2015.12.003
[arXiv:1510.00442 [nucl-th]].

\bibitem{ALICE:2016fzo}
J.~Adam \textit{et al.} [ALICE],
Nature Phys. \textbf{13} (2017), 535-539
doi:10.1038/nphys4111
[arXiv:1606.07424 [nucl-ex]].

\bibitem{Frixione:1997ma}
S.~Frixione, M.~L.~Mangano, P.~Nason and G.~Ridolfi,
Adv. Ser. Direct. High Energy Phys. \textbf{15} (1998), 609-706
doi:10.1142/9789812812667\_0009
[arXiv:hep-ph/9702287 [hep-ph]].

\bibitem{Norrbin:2000zc}
E.~Norrbin and T.~Sjostrand,
Eur. Phys. J. C \textbf{17} (2000), 137-161
doi:10.1007/s100520000460
[arXiv:hep-ph/0005110 [hep-ph]].

\bibitem{Braun-Munzinger:2003pwq}
P.~Braun-Munzinger, K.~Redlich and J.~Stachel,
doi:10.1142/9789812795533\_0008
[arXiv:nucl-th/0304013 [nucl-th]].

\bibitem{Alberico:2013bza}
W.~M.~Alberico, A.~Beraudo, A.~De Pace, A.~Molinari, M.~Monteno, M.~Nardi, F.~Prino and M.~Sitta,
Eur. Phys. J. C \textbf{73} (2013), 2481
doi:10.1140/epjc/s10052-013-2481-z
[arXiv:1305.7421 [hep-ph]].

\bibitem{Levai:2009mn}
P.~Levai and V.~Skokov,
Phys. Rev. D \textbf{82} (2010), 074014
doi:10.1103/PhysRevD.82.074014
[arXiv:0909.2323 [hep-ph]].

\bibitem{ALICE:2019oyn}
S.~Acharya \textit{et al.} [ALICE],
Eur. Phys. J. C \textbf{80} (2020) no.10, 979
doi:10.1140/epjc/s10052-020-8118-0
[arXiv:1910.14403 [nucl-ex]].

\bibitem{ALICE:2021kpy}
S.~Acharya \textit{et al.} [ALICE],
Eur. Phys. J. C \textbf{82} (2022) no.4, 335
doi:10.1140/epjc/s10052-022-10267-3
[arXiv:2110.10043 [nucl-ex]].


\bibitem{CMS:2020geg}
A.~M.~Sirunyan \textit{et al.} [CMS],
JHEP \textbf{05} (2021), 054
doi:10.1007/JHEP05(2021)054
[arXiv:2005.14219 [hep-ex]].

\bibitem{Beraudo:2014boa}
A.~Beraudo, A.~De Pace, M.~Monteno, M.~Nardi and F.~Prino,
Eur. Phys. J. C \textbf{75} (2015) no.3, 121
doi:10.1140/epjc/s10052-015-3336-6
[arXiv:1410.6082 [hep-ph]].

\bibitem{PHENIX:2018wex}
A.~Adare \textit{et al.} [PHENIX],
Phys. Rev. C \textbf{99} (2019) no.5, 054903
doi:10.1103/PhysRevC.99.054903
[arXiv:1803.01749 [hep-ex]].

\bibitem{Zhang:2019bkf}
L.~Y.~Zhang, J.~H.~Chen, Z.~W.~Lin, Y.~G.~Ma and S.~Zhang,
Phys. Rev. C \textbf{99} (2019) no.5, 054904
doi:10.1103/PhysRevC.99.054904
[arXiv:1904.08603 [nucl-th]].

\bibitem{Zhang:2018ucx}
L.~Y.~Zhang, J.~H.~Chen, Z.~W.~Lin, Y.~G.~Ma and S.~Zhang,
Phys. Rev. C \textbf{98} (2018) no.3, 034912
doi:10.1103/PhysRevC.98.034912
[arXiv:1808.10641 [nucl-th]].

\bibitem{PHENIX:2010hmo}
A.~Adare \textit{et al.} [PHENIX],
Phys. Rev. Lett. \textbf{107} (2011), 142301
doi:10.1103/PhysRevLett.107.142301
[arXiv:1010.1246 [nucl-ex]].

\bibitem{Eskola:2001gt}
K.~J.~Eskola, V.~J.~Kolhinen and R.~Vogt,
Nucl. Phys. A \textbf{696} (2001), 729-746
doi:10.1016/S0375-9474(01)01221-0
[arXiv:hep-ph/0104124 [hep-ph]].


\bibitem{Kharzeev:2005zr}
D.~Kharzeev and K.~Tuchin,
Nucl. Phys. A \textbf{770} (2006), 40-56
doi:10.1016/j.nuclphysa.2006.01.017
[arXiv:hep-ph/0510358 [hep-ph]].

\bibitem{ATLAS:2012cix}
G.~Aad \textit{et al.} [ATLAS],
Phys. Rev. Lett. \textbf{110} (2013) no.18, 182302
doi:10.1103/PhysRevLett.110.182302
[arXiv:1212.5198 [hep-ex]].

\bibitem{Albacete:2018ruq}
J.~L.~Albacete, G.~Giacalone, C.~Marquet and M.~Matas,
Phys. Rev. D \textbf{99} (2019) no.1, 014002
doi:10.1103/PhysRevD.99.014002
[arXiv:1805.05711 [hep-ph]].

\bibitem{Andersson:1983jt}
B.~Andersson, G.~Gustafson and B.~Soderberg,
Z. Phys. C \textbf{20} (1983), 317
doi:10.1007/BF01407824

\bibitem{Hwa:1979pn}
R.~C.~Hwa,
Phys. Rev. D \textbf{22} (1980), 1593
doi:10.1103/PhysRevD.22.1593

\bibitem{Buckley:2011ms}
A.~Buckley, J.~Butterworth, S.~Gieseke, D.~Grellscheid, S.~Hoche, H.~Hoeth, F.~Krauss, L.~Lonnblad, E.~Nurse and P.~Richardson, \textit{et al.}
Phys. Rept. \textbf{504} (2011), 145-233
doi:10.1016/j.physrep.2011.03.005
[arXiv:1101.2599 [hep-ph]].

\bibitem{Fujii:2013yja}
H.~Fujii and K.~Watanabe,
Nucl. Phys. A \textbf{920} (2013), 78-93
doi:10.1016/j.nuclphysa.2013.10.006
[arXiv:1308.1258 [hep-ph]].

\bibitem{ALICE:2018lyv}
S.~Acharya \textit{et al.} [ALICE],
JHEP \textbf{10} (2018), 174
doi:10.1007/JHEP10(2018)174
[arXiv:1804.09083 [nucl-ex]].

\bibitem{CMS:2017qjw}
A.~M.~Sirunyan \textit{et al.} [CMS],
Phys. Lett. B \textbf{782} (2018), 474-496
doi:10.1016/j.physletb.2018.05.074
[arXiv:1708.04962 [nucl-ex]].

\bibitem{ALICE:2015zhm}
J.~Adam \textit{et al.} [ALICE],
Phys. Lett. B \textbf{754}, 81-93 (2016)
doi:10.1016/j.physletb.2015.12.067
[arXiv:1509.07491 [nucl-ex]].


\bibitem{STAR:2014wif}
L.~Adamczyk \textit{et al.} [STAR],
Phys. Rev. Lett. \textbf{113} (2014) no.14, 142301
[erratum: Phys. Rev. Lett. \textbf{121} (2018) no.22, 229901]
doi:10.1103/PhysRevLett.113.142301
[arXiv:1404.6185 [nucl-ex]].

\bibitem{Connors:2017ptx}
M.~Connors, C.~Nattrass, R.~Reed and S.~Salur,
Rev. Mod. Phys. \textbf{90} (2018), 025005
doi:10.1103/RevModPhys.90.025005
[arXiv:1705.01974 [nucl-ex]].

\bibitem{Sjostrand:2006za}
T.~Sjostrand, S.~Mrenna and P.~Z.~Skands,
JHEP \textbf{05}, 026 (2006)
doi:10.1088/1126-6708/2006/05/026
[arXiv:hep-ph/0603175 [hep-ph]].

\bibitem{Singh:2021edu}
R.~Singh, Y.~Bailung and A.~Roy,
Phys. Rev. C \textbf{105}, no.3, 035202 (2022)
doi:10.1103/PhysRevC.105.035202
[arXiv:2108.08626 [nucl-th]].

\bibitem{Sjostrand:2007gs}
T.~Sjostrand, S.~Mrenna and P.~Z.~Skands,
Comput. Phys. Commun. \textbf{178}, 852-867 (2008)
doi:10.1016/j.cpc.2008.01.036
[arXiv:0710.3820 [hep-ph]].

\bibitem{Bierlich:2018xfw}
C.~Bierlich, G.~Gustafson, L.~L\"onnblad and H.~Shah,
JHEP \textbf{10} (2018), 134
doi:10.1007/JHEP10(2018)134
[arXiv:1806.10820 [hep-ph]].

\bibitem{ALICE:2016clc}
J.~Adam \textit{et al.} [ALICE],
Eur. Phys. J. C \textbf{77}, no.4, 245 (2017)
doi:10.1140/epjc/s10052-017-4779-8
[arXiv:1605.06963 [nucl-ex]].

\bibitem{Campbell:2022qmc}
J.~M.~Campbell, M.~Diefenthaler, T.~J.~Hobbs, S.~H\"oche, J.~Isaacson, F.~Kling, S.~Mrenna, J.~Reuter, S.~Alioli and J.~R.~Andersen, \textit{et al.}
[arXiv:2203.11110 [hep-ph]].

\bibitem{Biro:1994mp}
T.~S.~Biro, P.~Levai and J.~Zimanyi,
Phys. Lett. B \textbf{347} (1995), 6-12
doi:10.1016/0370-2693(95)00029-K

\bibitem{Fries:2008hs}
R.~J.~Fries, V.~Greco and P.~Sorensen,
Ann. Rev. Nucl. Part. Sci. \textbf{58} (2008), 177-205
doi:10.1146/annurev.nucl.58.110707.171134
[arXiv:0807.4939 [nucl-th]].

\bibitem{Kniehl:2007erq}
B.~A.~Kniehl, G.~Kramer, I.~Schienbein and H.~Spiesberger,
Phys. Rev. D \textbf{77} (2008), 014011
doi:10.1103/PhysRevD.77.014011
[arXiv:0705.4392 [hep-ph]].

\bibitem{Salajegheh:2019ach}
M.~Salajegheh, S.~M.~Moosavi Nejad, H.~Khanpour, B.~A.~Kniehl and M.~Soleymaninia,
Phys. Rev. D \textbf{99} (2019) no.11, 114001
doi:10.1103/PhysRevD.99.114001
[arXiv:1904.08718 [hep-ph]].

\bibitem{Epele:2018ewr}
M.~Epele, C.~Garc\'\i{}a Canal and R.~Sassot,
Phys. Lett. B \textbf{790} (2019), 102-107
doi:10.1016/j.physletb.2018.11.069
[arXiv:1807.07495 [hep-ph]].

\bibitem{Kramer:2017gct}
G.~Kramer and H.~Spiesberger,
Nucl. Phys. B \textbf{925} (2017), 415-430
doi:10.1016/j.nuclphysb.2017.10.016
[arXiv:1703.04754 [hep-ph]].

\bibitem{Kramer:2018vde}
G.~Kramer and H.~Spiesberger,
Phys. Rev. D \textbf{98} (2018) no.11, 114010
doi:10.1103/PhysRevD.98.114010
[arXiv:1809.04297 [hep-ph]].

\bibitem{Corke:2009tk}
R.~Corke and T.~Sjostrand,
JHEP \textbf{01} (2010), 035
doi:10.1007/JHEP01(2010)035
[arXiv:0911.1909 [hep-ph]].

\bibitem{Diehl:2011yj}
M.~Diehl, D.~Ostermeier and A.~Schafer,
JHEP \textbf{03} (2012), 089
[erratum: JHEP \textbf{03} (2016), 001]
doi:10.1007/JHEP03(2012)089
[arXiv:1111.0910 [hep-ph]].

\bibitem{Sjostrand:2017cdm}
T.~Sj\"ostrand,
Adv. Ser. Direct. High Energy Phys. \textbf{29} (2018), 191-225
doi:10.1142/9789813227767\_0010
[arXiv:1706.02166 [hep-ph]].

\bibitem{Andersson:1983ia}
B.~Andersson, G.~Gustafson, G.~Ingelman and T.~Sjostrand,
Phys. Rept. \textbf{97} (1983), 31-145
doi:10.1016/0370-1573(83)90080-7

\bibitem{Khoze:1994fu}
V.~A.~Khoze and T.~Sjostrand,
Phys. Lett. B \textbf{328} (1994), 466-476
doi:10.1016/0370-2693(94)91506-7
[arXiv:hep-ph/9403394 [hep-ph]].

\bibitem{Gieseke:2012ft}
S.~Gieseke, C.~Rohr and A.~Siodmok,
Eur. Phys. J. C \textbf{72} (2012), 2225
doi:10.1140/epjc/s10052-012-2225-5
[arXiv:1206.0041 [hep-ph]].

\bibitem{Rathsman:1998tp}
J.~Rathsman,
Phys. Lett. B \textbf{452} (1999), 364-371
doi:10.1016/S0370-2693(99)00291-9
[arXiv:hep-ph/9812423 [hep-ph]].

\bibitem{Corke:2010yf}
R.~Corke and T.~Sjostrand,
JHEP \textbf{03} (2011), 032
doi:10.1007/JHEP03(2011)032
[arXiv:1011.1759 [hep-ph]].

\bibitem{Bierlich:2016vgw}
C.~Bierlich, G.~Gustafson and L.~L\"onnblad,
[arXiv:1612.05132 [hep-ph]].

\bibitem{Bierlich:2017vhg}
C.~Bierlich, G.~Gustafson and L.~L\"onnblad,
Phys. Lett. B \textbf{779} (2018), 58-63
doi:10.1016/j.physletb.2018.01.069
[arXiv:1710.09725 [hep-ph]].

\bibitem{Bierlich:2014xba}
C.~Bierlich, G.~Gustafson, L.~L\"onnblad and A.~Tarasov,
JHEP \textbf{03} (2015), 148
doi:10.1007/JHEP03(2015)148
[arXiv:1412.6259 [hep-ph]].

\bibitem{Bierlich:2015rha}
C.~Bierlich and J.~R.~Christiansen,
Phys. Rev. D \textbf{92} (2015) no.9, 094010
doi:10.1103/PhysRevD.92.094010
[arXiv:1507.02091 [hep-ph]].



\bibitem{ALICE:2021dhb}
S.~Acharya \textit{et al.} [ALICE],
Phys. Rev. D \textbf{105}, no.1, L011103 (2022)
doi:10.1103/PhysRevD.105.L011103
[arXiv:2105.06335 [nucl-ex]].

\bibitem{Skands:2014pea}
P.~Skands, S.~Carrazza and J.~Rojo,
Eur. Phys. J. C \textbf{74}, no.8, 3024 (2014)
doi:10.1140/epjc/s10052-014-3024-y
[arXiv:1404.5630 [hep-ph]].

\bibitem{Christiansen:2015yqa}
J.~R.~Christiansen and P.~Z.~Skands,
JHEP \textbf{08}, 003 (2015)
doi:10.1007/JHEP08(2015)003
[arXiv:1505.01681 [hep-ph]].

\bibitem{Lonnblad:2021fyl}
L.~L\"onnblad,
Nucl. Phys. A \textbf{1005} (2021), 121873
doi:10.1016/j.nuclphysa.2020.121873

\end{thebibliography}

 \end{document}